\documentclass[a4paper,conference]{IEEEtran}

\usepackage{cite}

\usepackage{tikz}
\usetikzlibrary{arrows,snakes,shapes,positioning,arrows,decorations.markings,calc}
\usetikzlibrary{dsp,chains}
\usepackage{psfrag}
\usepackage{color}
\definecolor{green}{rgb}{0,0.4,0.05}
\definecolor{red}{rgb}{0.8,0,0}

\usepackage{amsbsy}
\usepackage{amsmath}
\usepackage{fixmath}
\usepackage{amssymb}
\usepackage{graphicx}
\usepackage{tikz} \usetikzlibrary{arrows,snakes,shapes,shapes.geometric,positioning,decorations.pathreplacing,arrows,decorations.markings,calc,automata,backgrounds,petri}
\usepackage{epstopdf}
\usepackage{epsfig}
\usepackage{psfrag}
\usepackage{pgfplots}
\usepackage{parskip}
\usepackage{pgfplots}
\pgfplotsset{compat=newest}
\pgfplotsset{plot coordinates/math parser=false}
\newlength\figureheight
\newlength\figurewidth
\pgfplotsset{every axis plot/.append style={line width=0.8pt}}

\usepackage{amsmath}

\DeclareMathAlphabet{\mathbit}{OML}{cmr}{bx}{it}
\DeclareMathAlphabet{\mathsf}{OT1}{cmss}{m}{n}
\DeclareMathAlphabet{\mathbsf}{OT1}{cmss}{bx}{n}
\newcommand{\bs}{\mathbit}

\newcommand{\of}[1]{\mkern-2mu\left(#1\right)}

\usepackage{comment}
\usepackage{algorithm}
\usepackage{algorithmic}

\begin{document}

\title{16 QAM Communication with 1-Bit Transmitters}
%
\author{\IEEEauthorblockN{Donia Ben Amor, Hela Jedda, Josef A. Nossek }
\IEEEauthorblockA{Associate Professorship of Signal Processing\\Technical University of Munich, 80290 Munich, Germany\\
Email: ga38pog@mytum.de;~\{hela.jedda;~josef.a.nossek\}@tum.de}
}

\maketitle
\begin{abstract}
We present a novel non-linear precoding technique for the transmission of 16 quadrature amplitude modulation (QAM) symbols in a 1-bit massive multi-user (MU) multiple-input-single-output (MISO) downlink system. We deploy low resolution digital-to-analog converters (DACs) at the transmitter for the sake of decreasing the high energy consumption related to the massive MISO system. To mitigate the multi-user interference (MUI) and the distortions due to the low resolution DACs, the minimum bit error ratio (MBER) precoder was introduced in previous work. However, this precoder technique is restricted to quadrature phase shift keying (QPSK) signaling. Our approach consists in upgrading this method to the transmission of 16 QAM symbols. Simulation results show that the performance in terms of uncoded BER is significantly improved for larger massive MISO gain.
\end{abstract}
\section{Introduction}
\label{sec:intro}

For the next generation of mobile communication energy efficiency is gaining more interest. We talk about green communication. One aspect of reducing the energy consumption of the mobile communication is the reduced energy consumption of the hardware mainly the power amplifiers at the transmitter side that are considered as the most power hungry devices at the transmitter side \cite{Blume2010, Chen2010}. When the power amplifiers are run in the saturation region high energy efficient systems are achieved. However, the saturation region is combined with high distortions and strong nonlinearities that are introduced to the signals. To avoid the PA distortions when run in the saturation region we resort to PA input signals of constant envelope. Constant envelope signals have the property of constant magnitude. Thus, the amplitude distortions are omitted.

The use of 1-bit digital-to-analog converters (DACs) at the transmitter ensures on the one hand the property of constant envelope signals at the input of the PA. On the other hand the energy consumption of the DAC itself is minimized. Therefore, the energy efficiency goal is achieved twice: energy efficient PA due to the constant envelope signals and energy efficient DACs due to the low resolution.

The contribution in \cite{Mezghani2009} is the first work that addressed the precoding task with low resolution quantization at the transmitter. The authors in \cite{Usman2016} introduced another linear precoder that could slightly improve the system performance. The proposed precoder is designed based on an iterative algorithm since no closed form expression can be obtained. Theoretical analysis on the achievable rate in systems with 1-bit transmitters were introduced in \cite{Kakkavas2016, Saxena2016, Yongzhi2016}. The first nonlinear precoding technique in this topic was presented in \cite{JeddaSAM2016}. The authors presented a symbol-wise precoding technique based on the so called minimum bit error ratio (MBER) criterion and made use of the box norm ($\ell_{\infty}$) to relax the 1-bit constraint. In \cite{Jacobsson_Studer2016_1} the authors present another symbol-wise precoder based on the minimum mean square error (MMSE) and extended it to higher modulation scheme in \cite{Jacobsson_Studer2016_2}. In this work we introduce another technique to transmit 16 QAM symbols despite of the 1-bit transmitters based on the idea of superposition coding. This method is based on symbol-wise optimization unlike the method in \cite{Jacobsson_Studer2016_2}, where the optimization is performed block-wise. This leads to lower complexity and thus lower latency. In addition, the receive filter design does not require any knowledge about the noise statistics.

This paper is organized as follows: in Section \ref{sec:sysmodel} we present the system model. In Sections \ref{sec:mapping} and \ref{sec:receiver} we review the MBER precoder that is restricted to QPSK signaling and introduce the extension to 16 QAM signals. A linear precoding technique from the literature is presented in Section \ref{sec:existing_precoder}. In Sections \ref{sec:sim} and \ref{sec:conclusion} we show the simulation results and summarize this work.

\textbf{Notation}: Bold letters indicate vectors and matrices, non-bold letters express scalars. The operators $(.)^{*}$, $(.)^{\rm T}$ and $(.)^{\rm H}$ stand for complex conjugation, the transposition and Hermitian transposition, respectively. The $n \times n$ identity (zeros) matrix is denoted by $\mathbf{I}_{n}$ ($\mathbf{0}_{n}$).
\section{System Model}
\label{sec:sysmodel}
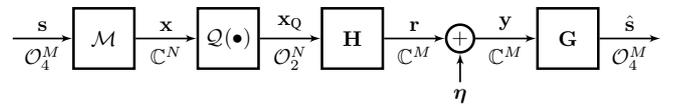
\begin{figure}[h]
\centering
\scalebox{.8}{\tikzstyle{int}=[draw, minimum width=1cm, minimum height=1cm,  very thick]
\tikzstyle{init} = [pin edge={<-,thick,black}]
\tikzstyle{sum} = [draw, circle,inner sep=1pt, minimum size=2mm, very thick] 

\begin{tikzpicture}[node distance=1cm,auto,>=latex']
    \node [int] (a) {$\mathcal{M}$};
    \node (b) [left of=a,node distance=1.5cm, coordinate] {a};
    \node [int] (c) [right=1cm of a] {$\mathcal{Q}\of{\bullet}$};
    \node [int] (e) [right=1cm of c] {$\mathbf{H}$};
    \node [sum,  pin={[init]below:$\bs{\eta}$}] (g) [right=1cm of e] {$\mathbf{+}$};
   
    \node [int] (i) [right=1cm of g] {$\mathbf{G}$};
      
    \node [coordinate] (end) [right=1cm of i, node distance=1cm]{};

    \path[->,thick] (b) edge node[above] {$\mathbf{s}$} 
                       node[below] {$\mathcal{O}_4^{M}$} (a);
    \path[->,thick] (a) edge node[above] {$\mathbf{x}$}
                       node[below] {$\mathbb{C}^{N}$} (c);
    \path[->,thick] (c) edge node[above] {$\mathbf{x}_\text{Q}$} 
                       node[below] {$\mathcal{O}_2^{N}$} (e);
    \path[->,thick] (e) edge node[above] {$\mathbf{r}$} 
                       node[below] {$\mathbb{C}^{M}$} (g);
    \path[->,thick] (g) edge node[above] {$\mathbf{y}$} 
                       node[below] {$\mathbb{C}^{M}$} (i);                      
    \path[->,thick] (i) edge node[above] {$\hat{\mathbf{s}}$} 
                      node[below] {$\mathcal{O}_4^{M}$} (end) ;
\end{tikzpicture}}
\caption{Downlink system model for 16 QAM symbols}
\label{fig:model}
\end{figure}
The system model shown in Fig.\ref{fig:model} consists of a massive MISO downlink scenario. The base station is equipped with $N$ antennas and serves $M$ single-antenna users simultaneously, where $N\gg M$.\\
If we consider $\mathcal{O}_4$ as the set of the 16 QAM constellation, the transmit signal $\mathbf{s}\in\mathcal{O}_4^M$ contains the symbols to be transmitted to each of the $M$ users. We assume that $\mathrm{E}[\mathbf{s}]=\mathbf{0}_{M}$ and $\mathrm{E}[\mathbf{s}\mathbf{s}^{\mathrm{H}}]=\sigma^2_\text{s}\mathbf{I}_{M}$. The signal vector $\mathbf{s}$ is mapped into the vector $\mathbf{x}$ prior to the DAC in order to reduce the distortions caused by the coarse quantization and the MU interference (MUI). The mapping $\mathcal{M}$ based on a LUT is upgraded from the method intoduced in \cite{JeddaSAM2016} applied for QPSK symbols.  The 1-bit quantization delivers then the signal vector $\mathbf{x}_{Q}\in\mathcal{O}_2^N$, where $\mathcal{O}_2$ is the set of QPSK constellation. At the receiver side, a receive filter $\mathbf{G}$ which is a real-valued diagonal matrix that ensures the normalization of the power of the received signal at each user, and hence scales the received 16 QAM constellation points to their right locations.\\
The received signal vector $\hat{\mathbf{s}}\in \mathcal{O}_4^M$ can be written as follows $\hat{\mathbf{s}}=\mathbf{G}\left(\sqrt{\frac{E_\text{tx}}{N}}\mathbf{H}\mathbf{x}_{Q}+\bs{\eta}\right)$. Here $E_\text{tx}$ denotes the total transmit power. In this model equal power allocation is performed, which means that each transmit antenna gets scaled with $\sqrt{\frac{E_\text{tx}}{N}}$. $\mathbf{H}$ is the channel matrix with the $\left(m,n\right)\text{th}$ element $h_\text{mn}$ being the zero-mean unit-variance channel tap between the transmit antenna and the receive antenna. $\bs{\eta} \sim \mathcal{C} \mathcal{N}\of{ \mathbf{0}_{M}, \mathbf{C}_\text{$\bs{\eta}$}=\mathbf{I}_{M}}$ denotes the vector of the additive white Gaussian noise (AWGN) components at the $M$ receive antennas.

\section{Mapping}
\label{sec:mapping}
Unlike other linear precoding techniques, in this model no explicit precoder is designed, but rather the transmit signal vector $\mathbf{x}$ for a given input $\mathbf{s}$ depending on the channel $\mathbf{H}$, while assuming full CSI. As mentioned before we upgrade the mapping method restricted to QPSK symbols to the transmission 16 QAM symbols. To this end we review shortly the minimum BER (MBER) precoding technique.
\subsection{MBER Criterion}
The mapping $\mathcal{M}$ introduced in \cite{JeddaSAM2016} consists of 3 steps. First, an optimization problem is solved for all possible input combinations from the QPSK constellaton. Then, the solutions of this problem are stored in a LUT of size $N \times 4^M$. The latter is updated in each coherence slot, i.e.\,for each channel. In a last step, the input vector $\mathbf{s}$ is mapped into the transmit vector $\mathbf{x}$ according to its index in the LUT. \\
The optimization criterion is here the MBER. In order to minimize the BER we require that the receive signal is in the same quadrant as the desired signal and far from the decision thresholds. Further the transmit signal $\mathbf{x}$ is constrained to have entries from the QPSK constellation in order to ensure a linear behavior between $\mathbf{x}$ and $\mathbf{x}_Q$ and thus avoid the loss of information due to the coarse quantization.

To keep the receive signal $r$ in the same quadrant as the input signal $s$, we maximize  $\Re\{r s^{*}\}$ and minimize the phase $|\phi|$, where $\Re\{r s^{*}\}$ is the projection of the vector $r$ on the desired vector $s$ and $\phi$ is the angle between $s$ and $r$ with $\phi\in ]-\frac{\pi}{4},\frac{\pi}{4}[$. Hence, the optimization problem for single-user case reads as the following 
\begin{align}
\max_{\mathbf{x}\in\mathcal{O}_2^N}\Re\{\left(r s^{*}\right)^2\}&= \max_{\mathbf{x}\in\mathcal{O}_2^N}|r|^2|s|^2\cos\of{2\phi}.
\label{eq:problem1}
\end{align}

Now, we reformulate the optimization problem for the MU case. The problem expressed in (\ref{eq:problem1}) is now applied for each user
\begin{align}
\max_{\mathbf{x}\in\mathcal{O}_2^N}\Re\{\left(r_{m} s_{m}^{*}\right)^2\}&= \max_{\mathbf{x}\in\mathcal{O}_2^N}|r_{m}|^2|s_{m}|^2\cos\of{2\phi_{m}}\nonumber\\ 
& \text{for }\, m=1,2,\dots,M.
\label{eq:problem2}
\end{align}

To jointly maximize the $M$ functions of (\ref{eq:problem2}), we write them in a diagonal matrix $\mathbf{P}$ given by
\begin{equation}
\mathbf{P}=\Re\{\mathrm{diag}\of{\mathbf{r}\mathbf{s}^{\mathrm{H}}}^2\},\nonumber\\
\end{equation}
where we stack the receive and input signals of each user in $\mathbf{r}$ and $\mathbf{s}$, respectively. \\
Hence, we get the $M$ cost functions as the diagonal entries of $\mathbf{P}$. To jointly maximize them, we compute their product given by the determinant of $\mathbf{P}$. This choice relies on the fact that the product is maximized, if all values have a considerable contribution.

In order to obtain a solution which coincides with the global maximum, the optimization problem should be convex. Here this is not the case, since the solution set of $\mathbf{x}=\sum_{n=1}^N x_n\mathbf{e}_n\in\mathcal{O}_2^N$ is non-convex ($\mathcal{O}_2$ is formed by four points in the space). Therefore we relax our constraint to the set $\mathcal{O}_2^{\prime N}$ given by $|\Re\{x_n\}|\leq \frac{1}{\sqrt{2}}$ and $|\Im\{x_n\}|\leq \frac{1}{\sqrt{2}}, \, \text{for}\, n=1,2,\dots,N$. The solution set of each entry in $\mathbf{x}$ is then a square and thus convex. The relaxed optimization problem is therefore given by
\begin{align}
&\max_{\mathbf{x}\in\mathcal{O}_2^{\prime N}}\mathrm{det}\of{\mathbf{P}}.
\label{eq:optrelax}
\end{align}

\subsection{Solution of the Optimization Problem}
To find a solution for the relaxed problem and fulfill the constraint, we resort to the gradient projection method. In each iteration $i$ of the algorithm we compute the new value of $\mathbf{x}$ according to
\begin{equation}
\mathbf{x}_{\left(i+1\right)}=\mathbf{x}_{\left(i\right)}+\mu\left(\frac{\partial\mathrm{det}\of{\mathbf{P}_{\left(i\right)}}}{\partial \mathbf{x}}\right),
\label{eq:gradient}
\end{equation}
then we project the obtained vector in the set $\mathcal{O}'^N_2$, where $\mathcal{O}'_2$ is the relaxed version of the set $\mathcal{O}_2$ and is given by the box formed from the QPSK points. 

\subsection{Split of 16 QAM Symbol in two QPSK Symbols}
Since the MBER precoding technique is restricted to QPSK symbols, we consider each 16 QAM symbol as the superposition of two QPSK symbols and apply the mapping approach for both QPSK symbols. The superposition is given by the following formula
\begin{equation}
\mathbf{s}=2\mathbf{\tilde{s}_1}+\mathbf{\tilde{s}_2}.
\end{equation}
For single user case and an input signal $s$, the symbol $\tilde{s}_1$ determines in which quadrant our input is situated, and the second symbol $\tilde{s}_2$ defines the deviation of $\tilde{s}_1$ in the right direction.

The question arises now: how can we include the factor 2 for the transmission of the vector $\mathbf{\tilde{s}_1}$?\\
Since we have equal power allocation, we can not scale the transmit antennas differently. Thus, we transmit the signal vector $\tilde{\mathbf{s}}_1$ across $\frac{2}{3}$ of the number of antennas $N$, which means with $\frac{2}{3}$ of the available transmit power, while the remaining transmit power is deployed for the transmission of $\tilde{\mathbf{s}}_2$ across the other $\frac{1}{3}$ of number of antennas. Obviously, the BS has the required knowledge about the combinations of QPSK symbols resulting in the 16 QAM symbols.

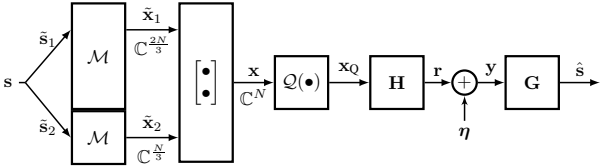
\begin{figure}[h]
\centering
\scalebox{.7}{\tikzstyle{int}=[draw, rectangle,minimum width=1cm, minimum height=1cm, very thick]
\tikzstyle{init} = [pin edge={<-,thick,black}]
\tikzstyle{sum} = [draw, circle,inner sep=1pt, minimum size=2mm, very thick] 

\begin{tikzpicture}[node distance=2cm,auto,>=latex']
    \node [int] (a) [minimum height=2cm]{$\mathcal{M}$};
    \node [int] (m) [below=0cm of a]{$\mathcal{M}$};
    \node [int] (l) [minimum height=3cm] at ([xshift=1.5cm, yshift=-0.5cm]a.east) {$\begin{bmatrix}
\bullet \\ \bullet  \end{bmatrix}$};
   
    \node (o) at ([xshift=-3.2cm]l.west) {$\mathbf{s}$};
    \node (p) at ([xshift=-3cm]l.west) {};
    \node [int] (c) [right=0.75cm of l] {$\mathcal{Q}({\bullet})$};
  
    \node [int] (e) [right=0.75cm of c] {$\mathbf{H}$};

    \node [sum,  pin={[init]below:$\bs{\eta}$}] (g) [right=0.5cm of e] {$\mathbf{+}$};
   
    \node [int] (i) [right=0.5cm of g] {$\mathbf{G}$};
      
    \node [coordinate] (end) [right=0.75cm of i, node distance=0.75cm]{};

   \draw[-,thick] (p.east) to node[midway] {} (o.east);  
   \draw[-latex,thick] (p.east) to node[midway,above=0.05cm]{$\tilde{\mathbf{s}}_1$} ([yshift=0.5cm]a.west);
   \draw[-latex,thick] (p.east) to node[midway,below=0.05cm]{$\tilde{\mathbf{s}}_2$} (m.west);
   \draw[-latex,thick] ([yshift=0.5cm]a.east) to node[midway,above]{$\tilde{\mathbf{x}}_1$}                     node[below]{$\mathbb{C}^{\frac{2N}{3}}$}([yshift=1cm]l.west);   
   \draw[-latex,thick] (m.east) to node[midway, above]{$\tilde{\mathbf{x}}_2$} node[below]{$\mathbb{C}^{\frac{N}{3}}$}([yshift=-1cm-1.2pt]l.west);    
   \draw[-latex,thick] (l.east) to node[midway,above]{$\mathbf{x}$} node[below]{$\mathbb{C}^N$} (c.west);
   \draw[-latex,thick] (c.east) to node[midway]{$\mathbf{x}_\text{Q}$} (e.west);            
   \draw[-latex,thick] (e.east) to node[midway]{$\mathbf{r}$} (g.west);
   \draw[-latex,thick] (g.east) to node[midway]{$\mathbf{y}$} (i.west); 
   \draw[-latex,thick] (i.east) to node[midway]{$\hat{\mathbf{s}}$} (end);
   
\end{tikzpicture}}
\caption{Downlink System Model with 16 QAM Symbols}
\label{fig:model2}
\end{figure}

At the transmitter side, we split the input vector $\mathbf{s}$ in the vectors $\tilde{\mathbf{s}}_1$ and $\tilde{\mathbf{s}}_2$ as depicted in Fig.\ref{fig:model2}.
Here we transmit $\tilde{\mathbf{s}}_1$ and $\tilde{\mathbf{s}}_2$, therefore we need to generate two LUTs, one of size $\frac{2N}{3}\times 4^M$ and the other has the size $\frac{N}{3}\times 4^M$. This means that we solve the optimization problem in (\ref{eq:optrelax}) for $\tilde{\mathbf{s}}_1$ and $\tilde{\mathbf{s}}_2$.\\
Since our optimization problem depends on the channel $\mathbf{H}$, we deploy for the mapping of $\tilde{\mathbf{s}}_1$ the submatrix $\tilde{\mathbf{H}}_1\in \mathbb{C}^{M\times \frac{2N}{3}}$. Whereas we use $\tilde{\mathbf{H}}_2\in\mathbb{C}^{\frac{N}{3}\times 1}$ for the mapping of $\tilde{\mathbf{s}}_2$, where 
\begin{equation}
\mathbf{H}=\begin{bmatrix}
\tilde{\mathbf{H}}_1 & \tilde{\mathbf{H}}_2
\end{bmatrix}.
\end{equation}
According to the index of $\tilde{\mathbf{s}}_1$ and $\tilde{\mathbf{s}}_2$ in their respective LUTs, we obtain the vectors $\tilde{\mathbf{x}}_1\in\mathbb{C}^{\frac{2N}{3}\times 1}$ and $\tilde{\mathbf{x}}_2\in\mathbb{C}^{\frac{N}{3}\times 1}$ containing the solutions of the optimization problems for the inputs $\tilde{\mathbf{s}}_1$ and $\tilde{\mathbf{s}}_2$. $\tilde{\mathbf{x}}_1$ and $\tilde{\mathbf{x}}_2$ are then stacked in the vector $\mathbf{x}$ to be transmitted so that $\mathbf{x}=\begin{bmatrix}\tilde{\mathbf{x}}^\mathrm{T}_1 & \tilde{\mathbf{x}}^\mathrm{T}_2 \end{bmatrix}^\mathrm{T}$.\\
The superposition is then performed in the air by multiplying with the channel matrix $\mathbf{H}$.

\section{Receive Filter}
\label{sec:receiver}
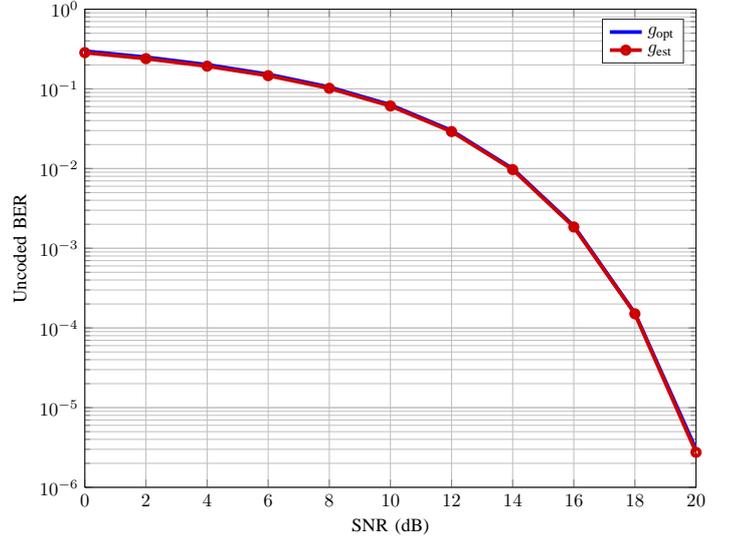
\begin{figure}[h!]
\centering
\scalebox{.7}{
%
%
\begin{tikzpicture}

\begin{axis}[%
width=4.521in,
height=3.566in,
at={(0.758in,0.481in)},
scale only axis,
xmin=0,
xmax=20,
xlabel={SNR (dB)},
xmajorgrids,
ymode=log,
ymin=1e-06,
ymax=1,
yminorticks=true,
ylabel={Uncoded BER},
ymajorgrids,
yminorgrids,
axis background/.style={fill=white},
legend style={legend cell align=left,align=left,draw=white!15!black}
]
\addplot [color=blue,solid,line width=2.0pt]
  table[row sep=crcr]{%
0	0.29772975\\
2	0.250093\\
4	0.2011845\\
6	0.1523515\\
8	0.1054005\\
10	0.0630385\\
12	0.0300065\\
14	0.0099615\\
16	0.00190375\\
18	0.0001535\\
20	3e-06\\
};
\addlegendentry{$g_{\text{opt}}$};

\addplot [color=red,solid,line width=2.0pt,mark=o,mark options={solid}]
  table[row sep=crcr]{%
0	0.28511425\\
2	0.23957975\\
4	0.1928325\\
6	0.1464855\\
8	0.10180975\\
10	0.06104775\\
12	0.02912725\\
14	0.009713\\
16	0.00185075\\
18	0.0001505\\
20	2.75e-06\\
};
\addlegendentry{$g_{\text{est}}$};

\end{axis}
\end{tikzpicture}%
}
\caption{Accuracy of the receive filter design given in (\ref{eq:g}) in SISO AWGN channel with 16 QAM signaling.}
\label{fig:g}
\end{figure}

In order to scale the expanded constellation obtained at the receiver, we include the receive filter 
\begin{align}
\mathbf{G}=\text{diag}\left( \begin{bmatrix}
g_1 & g_2 & \cdots &g_M
\end{bmatrix} \right).
\end{align} The scalar $g_m$ is used to fulfill that $\text{E}[\vert \Re\{\hat s_m\}\vert] = \text{E}[\vert \Re\{s_m\}\vert]$ and $\text{E}[\vert \Im\{\hat s_m\}\vert] = \text{E}[\vert \Im\{s_m\}\vert]$.
So, we compute
\begin{align}
\text{E}[\vert \Re\{s_m\}\vert] +\text{E}[\vert \Im\{s_m\}\vert]&=\text{E}[\vert \Re\{\hat s_m\}\vert] + \text{E}[\vert \Im\{\hat s_m\}\vert] \nonumber \\
&= g_m \text{E}[\vert \Re\{y_m\}\vert] + g_m \text{E}[\vert \Im\{y_m\}\vert] \nonumber \\
&= g_m \left( \text{E}[\vert \Re\{y_m\}\vert] +\text{E}[\vert \Im\{y_m\}\vert] \right),
\end{align}
which leads to
\begin{align}
g_m = \frac{\text{E}[\vert \Re\{s_m\}\vert ] + \text{E}[\vert \Im\{s_m\}\vert]}{\text{E}[\vert \Re\{y_m\}\vert ] + \text{E}[\vert \Im\{y_m\}\vert]}.
\label{eq:g}
\end{align}
Unlike the contribution in \cite{Jacobsson_Studer2016_2} the receive filter design does not require any knowledge about the noise. 
To show the accuracy of (\ref{eq:g}), we plot in Fig. \ref{fig:g} the uncoded BER as function of the signal-to-noise ratio (SNR) in a single-input-single-output (SISO) AWGN channel, where $\hat s = g (s+\eta)$ and $s \in \mathcal{O}_4$. The SNR is defined as $\text{SNR} =10 \log_{10}(\frac{\sigma_s^2}{\sigma_{\eta}^2})$. We compare the performance obtained with estimated $g_{\text{est}}$ according to (\ref{eq:g}) to the  performance obtained with the optimal $g_{\text{opt}} = \frac{1}{1+1/SNR}$.

\section{Existing Precoder}
\label{sec:existing_precoder}
In this section, we simulate the precoding scheme developed for the transmission of 16 QAM symbols, namely MBER with superposition in the air (MBER-sup). The results are compared to the linear precoder, named quantized Wiener filter (QWF) \cite{Mezghani2009}, which accounts for the distortions of the quantizer based on the MMSE criterion. The precoder is given by
\begin{align}
\mathbf{P}_{\text{QWF}}\!\!=\!\!\frac{1}{g_{\text{QWF}}}\!\!\left (\mathbf{H}^{\mathrm{H}}\mathbf{H}\!-\!\rho_{q}\textrm{nondiag} \left (\mathbf{H}^{\mathrm{H}}\mathbf{H}  \right ) \!\!+\frac{M\mathbf{I}_{N}} {E_{\textrm{tx}}}\right )^{-1}\!\!\!\mathbf{H}^{\mathrm{H}},
\end{align}
where 
\begin{align}
g_{\text{QWF}}&=\sqrt{\frac{\sigma_s^2 (1-\rho_q)}{E_{\textrm{tx}}}} \cdot \nonumber \\
&\sqrt{\textrm{tr}\left (\left (\mathbf{H}^\mathrm{H}\mathbf{H}\!-\!\rho_{q}\textrm{nondiag} \left (\mathbf{H}^\mathrm{H}\mathbf{H}  \right ) \!+\!\frac{M \mathbf{I}_{N}} {E_{\textrm{tx}}}\right )^{-2} \!\! \mathbf{H}^\mathrm{H}\mathbf{H} \right) },
\end{align}
and $\rho_q = 1-\frac{2}{\pi}$. 
In addition to the digital precoder $\mathbf{P}_{\text{QWF}}$, an analog precoder $\mathbf{D}$, given by a diagonal matrix, is included after the quantization operation in order to assign each antenna with a convenient amount of power, and minimize the quantization distortions of the 1-bit DAC. Thus, unequal power allocation is performed in each coherence slot.
\section{Simulation Results}
\label{sec:sim}
For our simulations, the performance metric is the uncoded BER. We consider i.i.d. Gaussian channel, as well as i.i.d. Gaussian noise with unit variance. The results are averaged over 100 channel realizations with $N_b=10^5$ transmit symbols per channel use and the used modulation scheme is 16 QAM.

\subsection{Comparison of MBER-sup and QWF}
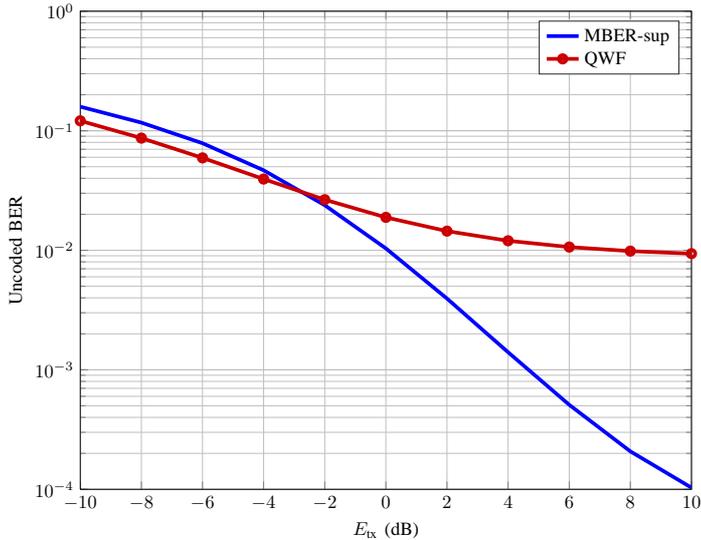
\begin{figure}
\centering
\scalebox{.7}{
%
%
\begin{tikzpicture}

\begin{axis}[%
width=4.521in,
height=3.566in,
at={(0.758in,0.481in)},
scale only axis,
xmin=-10,
xmax=10,
xlabel={$E_{\text{tx}}$ (dB)},
xmajorgrids,
ymode=log,
ymin=0.0001,
ymax=1,
yminorticks=true,
ylabel={Uncoded BER},
ymajorgrids,
yminorgrids,
axis background/.style={fill=white},
legend style={legend cell align=left,align=left,draw=white!15!black}
]
\addplot [color=blue,solid,line width=2.0pt]
  table[row sep=crcr]{%
-10	0.158946491666667\\
-8	0.116902041666667\\
-6	0.0786195083333333\\
-4	0.04669275\\
-2	0.0238188666666667\\
0	0.0103644583333333\\
2	0.00394818333333333\\
4	0.001402625\\
6	0.000509483333333333\\
8	0.00020785\\
10	0.000103266666666667\\
};
\addlegendentry{MBER-sup};

\addplot [color=red,solid,line width=2.0pt,mark=o,mark options={solid}]
  table[row sep=crcr]{%
-10	0.121056766666667\\
-8	0.0867703583333333\\
-6	0.0593019833333334\\
-4	0.039438475\\
-2	0.0265015416666667\\
0	0.018845775\\
2	0.014463525\\
4	0.0120129666666667\\
6	0.0106416666666667\\
8	0.00984199166666667\\
10	0.00937633333333334\\
};
\addlegendentry{QWF};

\end{axis}
\end{tikzpicture}%
}
\caption{Uncoded BER of the MBER-sup compared to QWF for a 1-bit system with $N=150$ transmit antennas and $M=3$ single-antenna users.}
\label{fig:plot1}
\end{figure}
We plot the uncoded BER as function of the transmit power $E_\text{tx}$ for a 1-bit MU-MISO system with $N=150$ transmit antennas and $M=3$ single-antenna users. Here we compare the precoder MBER-sup based on the superposition of two QPSK symbols in the air with the existing linear precoder QWF as shown in Fig.\ref{fig:plot1}.
It can be seen that the MBER-sup method outperforms the QWF precoder as well and reaches a BER of $10^{-3}$ at $E_\text{tx}$ value of almost $5$dB. The advantage of MBER-sup versus QWF consists in equal power allocation, i.e the signal processing part of the analog precoder is omitted and the available power is equally distributed among the transmit antennas, independently from the channel. Whereas, for QWF precoder, each antenna should be assigned with an appropriate amount of power. This is performed in each coherence slot.

\subsection{Performance while changing $M$ and $N$}
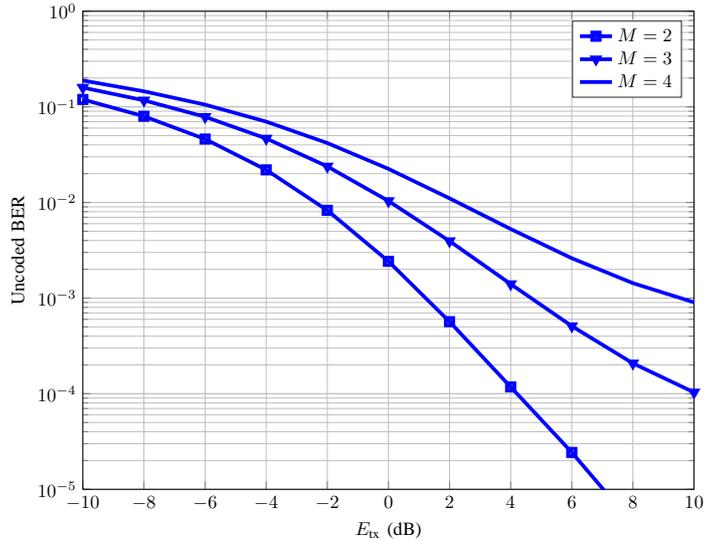
\begin{figure}[h!]
\centering
\scalebox{.7}{
%
%
\begin{tikzpicture}

\begin{axis}[%
width=4.521in,
height=3.566in,
at={(0.758in,0.481in)},
scale only axis,
xmin=-10,
xmax=10,
xlabel={$E_{\text{tx}}$ (dB)},
xmajorgrids,
ymode=log,
ymin=1e-05,
ymax=1,
yminorticks=true,
ylabel={Uncoded BER},
ymajorgrids,
yminorgrids,
axis background/.style={fill=white},
legend style={legend cell align=left,align=left,draw=white!15!black}
]
\addplot [color=blue,solid,line width=2.0pt,mark=square,mark options={solid}]
  table[row sep=crcr]{%
-10	0.1194370375\\
-8	0.0796530125\\
-6	0.0460335625\\
-4	0.0219343625\\
-2	0.00827787500000001\\
0	0.0024232625\\
2	0.0005666625\\
4	0.00011775\\
6	2.43125e-05\\
8	4.3375e-06\\
10	7.5e-07\\
};
\addlegendentry{$M=2$};

\addplot [color=blue,solid,line width=2.0pt,mark=triangle,mark options={solid,rotate=180}]
  table[row sep=crcr]{%
-10	0.158946491666667\\
-8	0.116902041666667\\
-6	0.0786195083333333\\
-4	0.04669275\\
-2	0.0238188666666667\\
0	0.0103644583333333\\
2	0.00394818333333333\\
4	0.001402625\\
6	0.000509483333333333\\
8	0.00020785\\
10	0.000103266666666667\\
};
\addlegendentry{$M=3$};

\addplot [color=blue,solid,line width=2.0pt]
  table[row sep=crcr]{%
-10	0.1885236375\\
-8	0.14524618125\\
-6	0.10536046875\\
-4	0.0700467625\\
-2	0.041836325\\
0	0.0224577375\\
2	0.01104796875\\
4	0.0052766875\\
6	0.0026115625\\
8	0.0014335\\
10	0.0009029375\\
};
\addlegendentry{$M=4$};

\end{axis}
\end{tikzpicture}%
}
\caption{BER of the MBER-sup precoder compared to QWF for a 1-bit system with $N=150$ transmit antennas and different number of single-antenna users.}
\label{fig:plot2}
\end{figure}

The performance of the system can be improved by increasing the ratio $\frac{N}{M}$, known as the massive MISO gain. In Fig.\ref{fig:plot2} the simulation results for the system with MBER-sup precoder are illustrated, while $N$ is kept constant to 150 transmit antennas and $M$ is varying. As expected, the performance for $M=4$ is the worst, since the ratio $\frac{N}{M}$ is the smallest in this case.

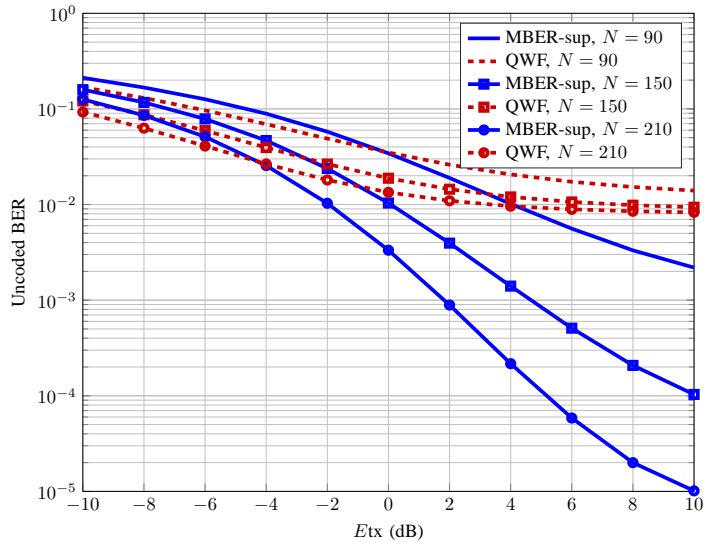
\begin{figure}[h!]
\centering
\scalebox{.7}{
%
%
\begin{tikzpicture}

\begin{axis}[%
width=4.521in,
height=3.566in,
at={(0.758in,0.481in)},
scale only axis,
xmin=-10,
xmax=10,
xlabel={$E{\text{tx}}$ (dB)},
xmajorgrids,
ymode=log,
ymin=1e-05,
ymax=1,
yminorticks=true,
ylabel={Uncoded BER},
ymajorgrids,
yminorgrids,
axis background/.style={fill=white},
title style={font=\bfseries},
legend style={legend cell align=left,align=left,draw=white!15!black}
]
\addplot [color=blue,solid,line width=2.0pt]
  table[row sep=crcr]{%
-10	0.211633175\\
-8	0.167352783333333\\
-6	0.12621085\\
-4	0.0891475916666666\\
-2	0.057787675\\
0	0.0342921583333333\\
2	0.0189597416666667\\
4	0.010166475\\
6	0.00559325\\
8	0.00331738333333333\\
10	0.00218956666666667\\
};
\addlegendentry{MBER-sup, $N=90$};

\addplot [color=red,dashed,line width=2.0pt]
  table[row sep=crcr]{%
-10	0.168746133333333\\
-8	0.130161591666667\\
-6	0.0965339416666667\\
-4	0.0692575083333333\\
-2	0.0489111166666667\\
0	0.0349930583333333\\
2	0.0260794666666667\\
4	0.020626175\\
6	0.0172871083333333\\
8	0.0152517833333333\\
10	0.0139892416666667\\
};
\addlegendentry{QWF, $N=90$};

\addplot [color=blue,solid,line width=2.0pt,mark=square,mark options={solid}]
  table[row sep=crcr]{%
-10	0.158946491666667\\
-8	0.116902041666667\\
-6	0.0786195083333333\\
-4	0.04669275\\
-2	0.0238188666666667\\
0	0.0103644583333333\\
2	0.00394818333333333\\
4	0.001402625\\
6	0.000509483333333333\\
8	0.00020785\\
10	0.000103266666666667\\
};
\addlegendentry{MBER-sup, $N=150$};

\addplot [color=red,dashed,line width=2.0pt,mark=square,mark options={solid}]
  table[row sep=crcr]{%
-10	0.121056766666667\\
-8	0.0867703583333333\\
-6	0.0593019833333334\\
-4	0.039438475\\
-2	0.0265015416666667\\
0	0.018845775\\
2	0.014463525\\
4	0.0120129666666667\\
6	0.0106416666666667\\
8	0.00984199166666667\\
10	0.00937633333333334\\
};
\addlegendentry{QWF, $N=150$};

\addplot [color=blue,solid,line width=2.0pt,mark=o,mark options={solid}]
  table[row sep=crcr]{%
-10	0.125833883333333\\
-8	0.0856240833333333\\
-6	0.051083875\\
-4	0.025560775\\
-2	0.0103088083333333\\
0	0.00332825833333333\\
2	0.000889633333333333\\
4	0.000216641666666667\\
6	5.855e-05\\
8	1.99833333333333e-05\\
10	1.01333333333333e-05\\
};
\addlegendentry{MBER-sup, $N=210$};

\addplot [color=red,dashed,line width=2.0pt,mark=o,mark options={solid}]
  table[row sep=crcr]{%
-10	0.092819\\
-8	0.0629179166666667\\
-6	0.0409057416666667\\
-4	0.0265342166666667\\
-2	0.0180595\\
0	0.01341105\\
2	0.010942975\\
4	0.00960925833333334\\
6	0.0089088\\
8	0.008505125\\
10	0.008288975\\
};
\addlegendentry{QWF, $N=210$};
\end{axis}
\end{tikzpicture}%
}
\caption{BER of the MBER-sup precoder compared to QWF for a 1-bit system with $M=3$ single-antenna users and different number of transmit antennas.}
\label{fig:plot3}
\end{figure}

Fig. \ref{fig:plot3} shows that increasing the number of transmit antennas does not improve the performance when using the QWF precoder. Whereas the performance can be increased significantly with MBER-sup.

\section{CONCLUSION}
\label{sec:conclusion}
Assuming full knowledge of CSI at the BS, we design a non-linear precoder for 1-bit MU-MISO downlink system with 16 QAM signaling.
Considering a system with single-antenna users, we deploy an approach based on the MBER precoding technique for a system transmitting QPSK symbols, where a sort of mapping of the input signal into the transmit signal is performed using a LUT. 
For our design called MBER-sup, we split the 16 QAM entries of the input vector into two QPSK symbols, deploy the same mapping technique for both QPSK symbols, then we obtain again the 16 QAM symbol through a superposition in the air. The latter is ensured by transmitting one QPSK symbol with two thirds of the available energy and the other symbol with the remaining other third of energy, where equal power allocation at the transmit antennas is performed.\\
The so-called MBER-sup approach outperforms the existing linear precoder QWF based on the MMSE criterion.
We also study the effect of the ratio $\frac{N}{M}$ on the performance of the system, and find that the bigger this ratio, also named as massive MISO gain, the lower is the BER and thus the better is the system performance.

\bibliographystyle{IEEEtran}
\bibliography{IEEEabrv,refs}
\end{document}